# Radio-Frequency Quantum Rectification in Kagome Superconductor $CsV_3Sb_5$

Han-Xin Lou[1*], Jing-Jing Chen[2*], Xing-Guo Ye[1*], Zhen-Bing Tan[2†], An-Qi Wang[1†],

Qing Yin[1], Xin Liao[1], Jing-Zhi Fang[1], Xing-Yu Liu[1], Yi-Lin He[1], Zhen-Tao Zhang[1],

Chuan Li[3], Zhong-Ming Wei[4], Xiu-Mei Ma[5], Dapeng Yu[2,6], Zhi-Min Liao[1,7†]

[1]State Key Laboratory for Mesoscopic Physics and Frontiers Science Center for Nano-optoelectronics, School of Physics, Peking University, Beijing 100871, China.
[2]International Quantum Academy, Shenzhen, Guangdong, China.
[3]MESA+ Institute for Nanotechnology, University of Twente, 7500 AE, Enschede, The Netherlands.
[4]State Key Laboratory of Semiconductor Physics and Chip Technologies, Institute of Semiconductors, Chinese Academy of Sciences, Beijing 100083, China.
[5]Electron Microscopy Laboratory, School of Physics, Peking University, Beijing 100871, China.
[6]Shenzhen Branch, Hefei National Laboratory, Shenzhen, China
[7]Hefei National Laboratory, Hefei 230088, China.

*These authors contributed equally to this work.
†Corresponding authors, E-mail: tanzhenbin@iqasz.cn, anqi0112@pku.edu.cn, liaozm@pku.edu.cn

**Abstract**

Rectification of electromagnetic fields into direct current (DC) is pivotal for energy harvesting, wireless charging, and next-generation communication technologies. The superconducting diode effect, which exploits the nonreciprocal transport of dissipationless superconducting currents, offers ultra-low power consumption and high rectification ratios. Combining the superconducting diode effect with the AC Josephson effect holds promise for converting radio-frequency (*rf*) irradiation into a quantized DC output. However, experimental realization has been hindered by challenges in achieving the necessary symmetry breaking and fabricating high-performance Josephson junctions. Here we demonstrate the quantum rectification in kagome superconductor $CsV_3Sb_5$, which hosts emergent Josephson effects and a zero-field Josephson diode. Under *rf* irradiation, a DC voltage emerges without applied bias, scaling linearly with frequency as $V = hf/2e$, where $h$ is Planck's constant, $f$ is the microwave frequency, and $e$ is the electron charge. Furthermore, the rectified voltage exhibits quantized steps with increasing *rf* power, consistent with Shapiro step quantization. Our work establishes $CsV_3Sb_5$ as a versatile platform for wireless quantum power supplies and charging, and underscores the intertwined order parameters as a promising pathway for precise quantum matter control.

**Main**

Symmetry breaking in superconductivity can lead to exotic phenomena. As a typical example, the superconducting diodes with nonreciprocal supercurrent lay the foundations towards ultralow-dissipation rectification beyond the p-n junction diodes[1–12]. Moreover, the realization of superconducting diodes by constructing Josephson junctions using superconductor/non-superconductor interfaces can break the spatial inversion symmetry or time-reversal symmetry, allowing for the development of controllable and scalable rectifiers[13–22]. The recently discovered vanadium-based kagome superconductors $AV_3Sb_5$ (A = K, Rb, or Cs) intrinsically host symmetry-breaking orders and intertwined quantum phases[23–30], offering an attractive platform for exploring Josephson diode functionalities. In $AV_3Sb_5$, the interplay among superconductivity[24], charge, spin, and pair density waves[31–35], as well as electronic nematicity[36,37] and a pronounced anomalous Hall effect[38], not only gives rise to unconventional superconductivity and nontrivial quantum phenomena [39–45], but also facilitates precise quantum state control for functional quantum devices.

Recent experimental studies have provided compelling evidence for the coexistence of intertwined orders in $CsV_3Sb_5$. Optical Kerr measurements[46] and angle-resolved photoemission spectroscopy[47] suggest the possible nodeless superconductivity in this system[48]. Moreover, the superconducting diode effect, marked by asymmetry between the positive and negative critical currents, has been observed in $CsV_3Sb_5$[49], along with superconducting domains evidenced by oscillations of the critical current in response to the magnetic field. Additional observations of magnetic-field-switchable chiral charge order[40] and chiral pair density waves[34,43,44] further suggest the chiral nature of these domains, supported by loop current theories[50,51]. Intriguingly, inherent weak links between superconducting domains in $CsV_3Sb_5$ may naturally give rise to Josephson coupling, rendering the system promising for the realization of Josephson diode effects.

In this work, we demonstrate the emergent Josephson effect and the rectification from *rf*-microwave to $V_{dc}$ output in a monolithic $CsV_3Sb_5$ sample. Unlike the conventional AC Josephson effect, where non-zero Shapiro voltage steps appear only with an applied DC current[52,53], the $CsV_3Sb_5$-based Josephson diode shows quantized Shapiro voltage steps under *rf* irradiation without applying any current bias. Functioning as a wireless rectifier, this *rf*-modulated Josephson diode holds promise for wireless power, information processing, sensing, and measurement standardization in quantum device applications[54–56].

Three devices, labeled Devices 1–3, were fabricated for the measurements. The CsV$_3$Sb$_5$ crystals used exhibit high quality, as confirmed by scanning transmission electron microscopy (Extended Data Fig. 1). The chosen CsV$_3$Sb$_5$ film of Device 1 has a thickness ~100 nm and a width ~1 μm (see Methods for fabrication details). Au electrodes with various channel lengths were fabricated, labeled as 1-5 in Device 1 (Fig. 1a); all data in the main text were obtained between electrodes 2-3 in Device 1, with a channel length of ~600 nm. Figure 1b shows the temperature-dependent resistance of the CsV$_3$Sb$_5$ film, measured via the four-probe method, revealing a clear superconducting transition at ~3 K, marked by a sharp drop in resistance. Figure 1c illustrates the effect of an out-of-plane magnetic field $B_z$ on the superconducting state, with a critical magnetic field $B_c$~1 T.

The differential resistance $dV/dI$ as a function of direct current $I_{dc}$ at zero magnetic field is depicted in Fig. 1d, where $I_{dc}$ is swept from zero to positive and negative values, respectively. The curves reveal an asymmetry in critical currents: $I_c^+$ in the positive direction is larger than $|I_c^-|$ in the negative direction, indicating a non-reciprocal transition characteristic analogous to p-n junction diodes[49,57]. Notably, this effect is significantly enhanced by applying a small magnetic field (see Supplemental Note 1 for details). Additionally, as shown in Fig. 1e, several distinct superconducting transition kinks (highlighted by color blocks and indicated by color arrows) appear in the $I-V$ curve, with the first kink in the orange block corresponding to the transition peak in Fig. 1d (see Supplemental Note 1 for further discussion).

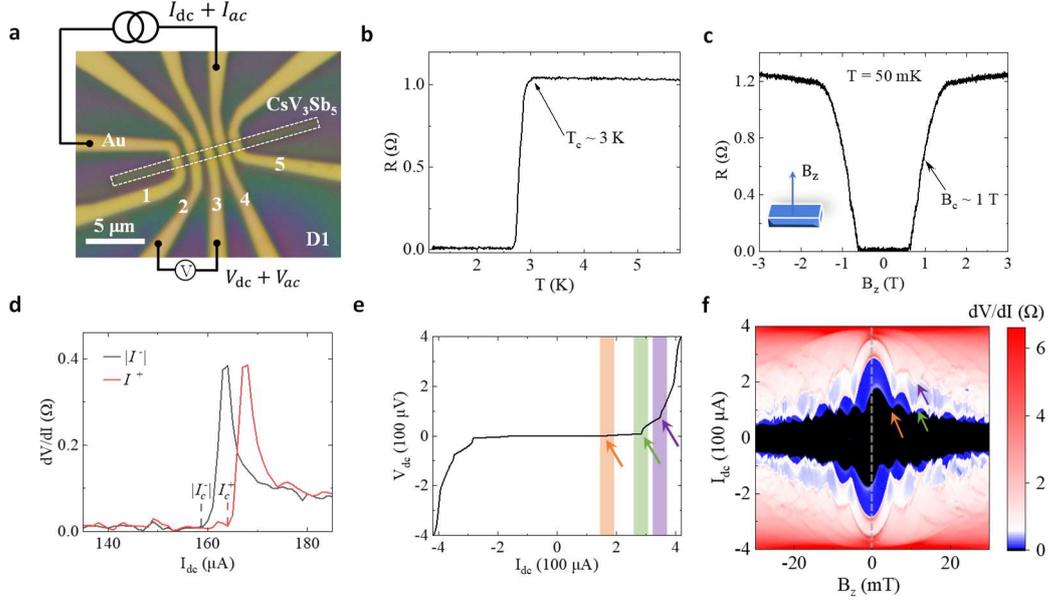

**Fig. 1: Superconducting characteristics of CsV₃Sb₅.**

**a**, Optical image of a typical CsV$_3$Sb$_5$ device. **b**, Temperature-dependent resistance measured by four-probe method, showing $T_c \sim 3$ K. **c**, Resistance as a function of out-of-plane magnetic field at 50 mK, with $B_c \sim 1$ T. **d**, $dV/dI$ as a function of $I_{dc}$ at $B = 0$ T and $T = 50$ mK. Black and red curves represent measurements with $I^-$ ($0 \rightarrow -200$ μA) and $I^+$ ($0 \rightarrow 200$ μA), respectively, indicating the zero-field diode effect. **e**, $I$-$V$ curve at $B = 0$ T and $T = 50$ mK, presenting multiple superconducting transitions, denoted by different color arrows. **f**, Color map of $dV/dI$ as a function of $I_{dc}$ and $B_z$ at $T = 50$ mK.

The $dV/dI$ measured at 50 mK is plotted against $I_{dc}$ and $B_z$ in Fig. 1f. The critical current $I_c$ exhibits a Fraunhofer-like interference pattern with oscillations as a function of $B_z$. Notably, three distinct oscillation modes with different decay rates can be clearly identified, as indicated by the arrows in Fig. 1f. These multiple supercurrent interference patterns suggest the presence of multiple supercurrent channels with different phases, likely associated with multiple self-assembled Josephson junctions within the CsV$_3$Sb$_5$ device[49]. Moreover, it is found that these interference patterns can be modulated by thermal cycles (Extended Data Fig. 2). A plausible explanation for the observed phenomena is the existence of dynamic superconducting domains[49] (see Supplemental Note 2 for details).

The interference pattern exhibits asymmetry between positive and negative currents across the magnetic field range, especially for the innermost mode (Fig. 1f). The central peak in the innermost $I_c - B_z$ pattern deviates from zero field, with the direction of deviation controlled by $I_{dc}$: positive $I_{dc}$ shifts the interference pattern towards positive $B_z$, while negative $I_{dc}$ shifts it in the negative direction. This shift of the peak center in $B_z$ is highly reproducible during repeated scans of the magnetic field in both positive and negative directions (Extended Data Fig. 3), indicating that it is not caused by vortex trapping during the magnetic field sweeping. The asymmetric pattern of the critical currents suggests the inversion symmetry breaking in $CsV_3Sb_5$[30,34,36,58].

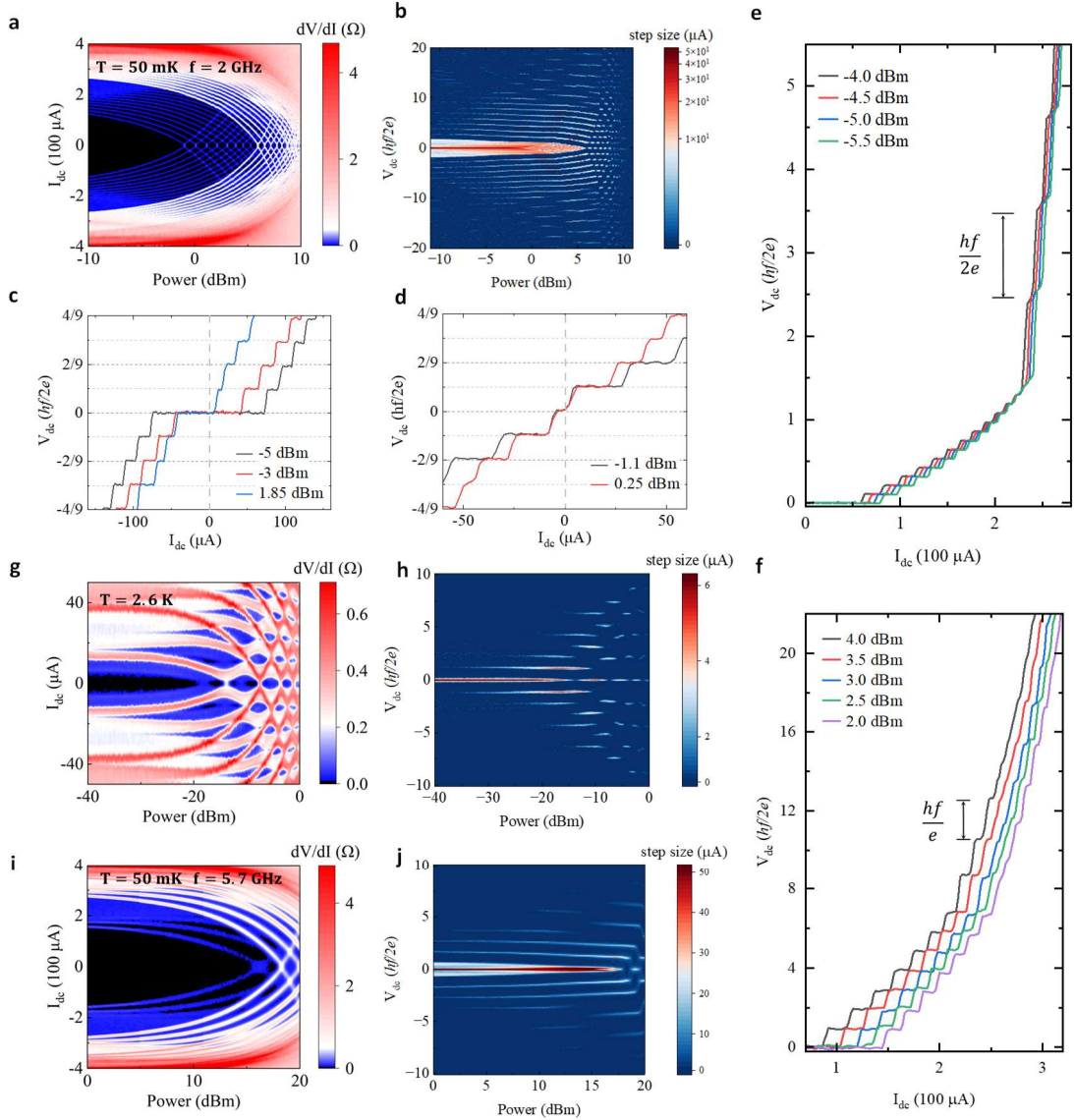

**Fig. 2: Emergent AC Josephson effect in monolithic CsV$_3$Sb$_5$ sample.**
**a-f**, The Josephson effect measured at $T = 50$ mK, $B = 0$ T, and $f = 2$ GHz. **a,** Color map of $dV/dI$ as a function of $I_{dc}$ and *rf* power; **b,** Mapping of step size as a function of $V_{dc}$ and *rf* power. **c-d,** $I - V$ curves at low $I_{dc}$ with different *rf* powers, presenting the Shapiro steps and *rf*-modulated Josephson diode effect. **e-f,** $I - V$ curves at different *rf* powers, presenting the evolution of different sets of Shapiro steps. **g-h,** The AC Josephson effect measured at higher temperature of $T = 2.6$ K, $B = 0$ T, and $f = 2$ GHz. **i-j,** The AC Josephson effect measured at higher frequency of $f = 5.7$ GHz, $T = 50$ mK and $B = 0$ T.

To further reveal the essence of the interference pattern, *rf* measurements were performed on the CsV$_3$Sb$_5$ junction. The device was irradiated by microwaves via a coaxial line, where the *rf* microwaves combine with an applied direct current, inducing Shapiro steps[52]. Figure 2a displays two-dimensional mapping of $dV/dI$ as a function of $I_{dc}$ and *rf* power ($P$), measured at 50 mK with frequency $f = 2$ GHz in the absence of a magnetic field. The clear superconducting interference pattern reveals oscillations of Shapiro steps, appearing sequentially one by one. Figure 2b shows the Shapiro step size at different voltages $V_{dc}$ and *rf* powers, where Shapiro steps at $V_n = n\frac{hf}{2e}$ are distinctly observed under microwave irradiation. The observed Shapiro steps directly indicate local Josephson junctions in the monolithic CsV$_3$Sb$_5$ sample, likely formed by weak links between superconducting domains[59]. Theoretically, such domain structures can emerge in multiband superconductors with competing order parameters, as in CsV$_3$Sb$_5$[60,61]. Domain walls act as natural weak links, mediating Josephson coupling and giving rise to the Fraunhofer-like interference and Shapiro steps. Their intrinsic inversion asymmetry and dynamics further lead to diode-like effects and thermally modulated interference patterns (see Supplemental Note 2). Alternative explanations, such as sample inhomogeneity and phase-slip lines, are discussed in Supplemental Note 3 but do not fully account for the data.

Figures 2c and 2d show the $I - V$ curves of the first set of Shapiro steps under different *rf* powers, exhibiting the fractional Shapiro steps (see more details about the fractional steps[20] in Supplemental Notes 4 and 5). It is noted that the $n = 0$ plateau in the $I - V$ curve exhibits the obvious Josephson diode effect, which is tunable by the *rf* powers as demonstrated by Fig. 2c. Furthermore, at certain *rf* powers, the $I - V$ curves display perfect non-reciprocal transmission for the $n = 0$ state, representing the ideal *rf*-modulated Josephson diode effect [4,62,63] (Fig. 2d). Compared to Fig. 1d, the diode effect is significantly enhanced under *rf* irradiation. We define the Josephson diode efficiency at the zeroth Shapiro step ($V_{dc} = 0$) as $\tilde{\eta}$. Under specific *rf* irradiation, $\tilde{\eta}$ can reach 100%, meaning that the zeroth step is entirely located on one side of $I_{dc} = 0$ (see Supplemental Note 6 for further discussions about the *rf*-modulated Josephson diode effect). This indicates that the device can generate a nonzero $V_{dc}$ output solely from *rf* irradiation, even when $I_{dc} = 0$, thereby showcasing the *rf* rectification effect. Furthermore, the application of small magnetic fields of ±2.5 mT significantly enhances the *rf* rectification effect even at relatively low *rf* power conditions (Supplemental Fig. S15).

Various sets of Shapiro steps are identified in Fig. 2a, each of which corresponds to a superconducting transition in Fig. 1e (see Methods and Extended Data Figs. 4-5 for detailed discussions). Figures 2e and 2f present the evolution of different sets of Shapiro steps under various *rf* powers. With $I_{dc}$ increases, the Shapiro steps transition from fractional to integer values with step height of $\frac{hf}{2e}$, and then evolve to higher-order step with step height of $\frac{hf}{e}$, where the latter is attributed to two Josephson junctions in series.

Figure 2g presents *rf* measurements at a higher temperature of 2.6 K, near the superconducting transition temperature of $CsV_3Sb_5$. The Shapiro steps remain visible. The distinct Shapiro steps at high temperatures close to $T_c$ highlights the robustness of the AC Josephson effect in this device. Additionally, the AC Josephson effect also persists at higher frequency. Figure 2i shows the $dV/dI$ as a function of $I_{dc}$ and *rf* power at a frequency of $f = 5.7$ GHz and at 50 mK. It's found that the integer Shapiro steps remain prominent regardless of the temperature and *rf* frequency. See more details about temperature dependence of Josephson effect in Extended Data Fig. 6. Furthermore, the superconducting interference patterns and AC Josephson effect were examined in different channels of several devices (D1-D3), all showing reproducible results (Supplemental Notes 7 and 8).

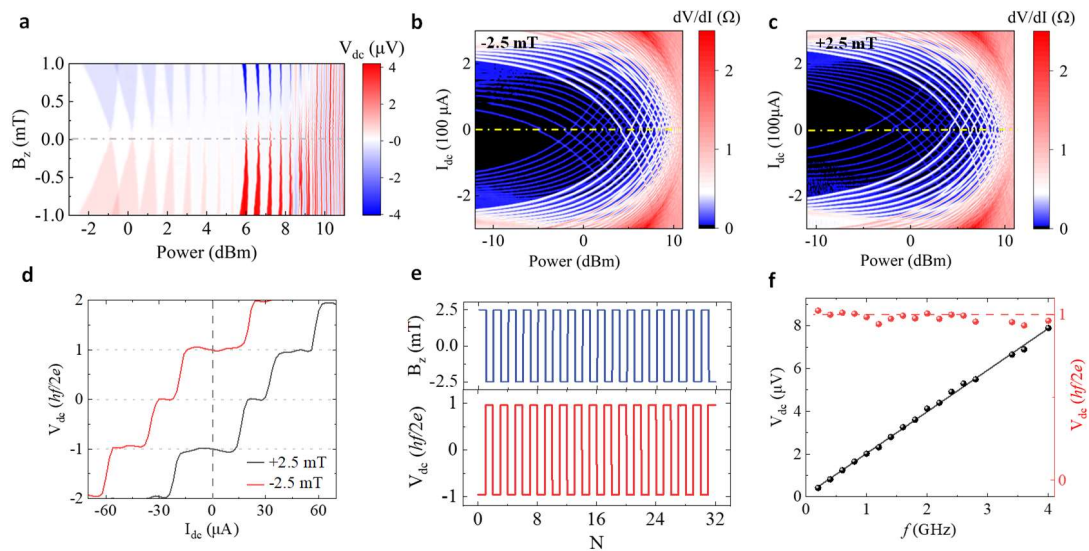

**Fig. 3: Radio-frequency rectification effect of the Josephson diode.**
**a**, Mapping of $V_{dc}$ output at $I_{dc} = 0$ as a function of magnetic field and *rf* power. **b** and **c**, Mapping of $dV/dI$ as a function of $I_{dc}$ and *rf* power with magnetic field of (**b**) $-2.5$ mT and (**c**) $2.5$ mT, respectively. **d**, $I - V$ curves at *rf* power of 4.9 dBm with ±2.5 mT. **e,** Under $I_{dc} = 0$ and *rf* power of 4.9 dBm, the output $V_{dc}$ switches between $\mp hf/2e$ as the magnetic field toggles between ±2.5 mT. **f**, Frequency dependence of the output DC voltage under $I_{dc} = 0$ and $B_z = -2.5$ mT, consistent with the relation of $V_{dc}$ = $hf/2e$, where the quantized $V_{dc}$ at each frequency requires a specific *rf* power to achieve. Data were acquired at $T = 50$ mK, and the microwave frequency in **a** – **e** is set at $f = 2$ GHz.

The high-frequency rectification capability of the $CsV_3Sb_5$-based device has been further demonstrated. Figure 3a presents the output $V_{dc}$ as a function of the out-of-

10 / 39power at a frequency of $f = 5.7$ GHz and at 50 mK. It's found that the integer Shapiro steps remain prominent regardless of the temperature and *rf* frequency. See more details about temperature dependence of Josephson effect in Extended Data Fig. 6. Furthermore, the superconducting interference patterns and AC Josephson effect were examined in different channels of several devices (D1-D3), all showing reproducible results (Supplemental Notes 7 and 8).

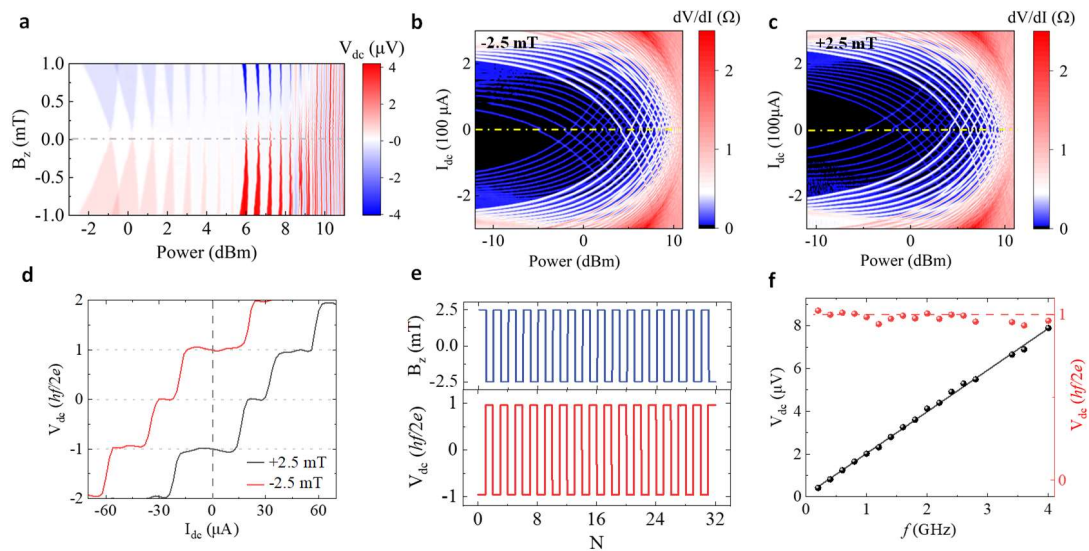

**Fig. 3: Radio-frequency rectification effect of the Josephson diode.**
**a**, Mapping of $V_{dc}$ output at $I_{dc} = 0$ as a function of magnetic field and *rf* power. **b** and **c**, Mapping of $dV/dI$ as a function of $I_{dc}$ and *rf* power with magnetic field of (**b**) $-2.5$ mT and (**c**) $2.5$ mT, respectively. **d**, $I - V$ curves at *rf* power of 4.9 dBm with ±2.5 mT. **e,** Under $I_{dc} = 0$ and *rf* power of 4.9 dBm, the output $V_{dc}$ switches between $\mp hf/2e$ as the magnetic field toggles between ±2.5 mT. **f**, Frequency dependence of the output DC voltage under $I_{dc} = 0$ and $B_z = -2.5$ mT, consistent with the relation of $V_{dc}$ = $hf/2e$, where the quantized $V_{dc}$ at each frequency requires a specific *rf* power to achieve. Data were acquired at $T = 50$ mK, and the microwave frequency in **a** – **e** is set at $f = 2$ GHz.

The high-frequency rectification capability of the $CsV_3Sb_5$-based device has been further demonstrated. Figure 3a presents the output $V_{dc}$ as a function of the out-of-plane magnetic field and *rf* power at $I_{dc} = 0$ and $f = 2$ GHz. A finite $V_{dc}$ is clearly observed under the application of a magnetic field. Moreover, as the *rf* power increases, a finite $V_{dc}$ can emerge even in the absence of both a magnetic field and a DC current bias, as indicated by the dashed line at $B_z = 0$ in Fig. 3a.

This $V_{dc}$ generated at $I_{dc} = 0$ is further confirmed through measurements of the AC Josephson effect under $B_z = \pm 2.5$ mT. As shown in Figs. 3b and 3c, the $dV/dI$ maps as a function of *rf* power and $I_{dc}$ reveal a pronounced asymmetry around zero $I_{dc}$ over the entire *rf* power range. The yellow dashed lines at $I_{dc} = 0$ in Figs. 3b and 3c indicate the emergence of nonzero DC voltages at specific *rf* powers (Extended Data Fig. 7). The typical $I - V$ characteristics at *rf* power 4.9 dBm are depicted in Fig. 3d, where the zeroth voltage step is fully positioned on one side of $I_{dc} = 0$. Clearly, the $V_{dc}$ exhibits nonzero output at $I_{dc} = 0$, with the voltage value exactly at the first Shapiro step, demonstrating the rectification from the *rf*-microwave to a quantized voltage. At $I_{dc} = 0$, the $V_{dc}$ output exhibits opposite Shapiro steps when the magnetic field is reversed. The modulation of the *rf* rectification effect by the magnetic field is further investigated with the $V_{dc}$ map plotted as a function of $I_{dc}$ and $B_z$ under a reduced *rf* power of 2.6 dBm (see Extended Data Fig. 7e). This map clearly shows a nonzero $V_{dc}$ at $I_{dc} = 0$ and a sign reversal when the magnetic field direction is flipped. The polarity reversal of the *rf* rectification through magnetic field flipping is further demonstrated by alternating between ±2.5 mT magnetic fields at $I_{dc} = 0$, as shown in Fig. 3e, where the output $V_{dc}$ continuously toggles between the "-h*f*/2e" and "h*f*/2e" states. Moreover, these quantized output voltages under *rf* irradiation pulses show

11 / 39plane magnetic field and *rf* power at $I_{dc} = 0$ and $f = 2$ GHz. A finite $V_{dc}$ is clearly observed under the application of a magnetic field. Moreover, as the *rf* power increases, a finite $V_{dc}$ can emerge even in the absence of both a magnetic field and a DC current bias, as indicated by the dashed line at $B_z = 0$ in Fig. 3a.

This $V_{dc}$ generated at $I_{dc} = 0$ is further confirmed through measurements of the AC Josephson effect under $B_z = \pm 2.5$ mT. As shown in Figs. 3b and 3c, the $dV/dI$ maps as a function of *rf* power and $I_{dc}$ reveal a pronounced asymmetry around zero $I_{dc}$ over the entire *rf* power range. The yellow dashed lines at $I_{dc} = 0$ in Figs. 3b and 3c indicate the emergence of nonzero DC voltages at specific *rf* powers (Extended Data Fig. 7). The typical $I - V$ characteristics at *rf* power 4.9 dBm are depicted in Fig. 3d, where the zeroth voltage step is fully positioned on one side of $I_{dc} = 0$. Clearly, the $V_{dc}$ exhibits nonzero output at $I_{dc} = 0$, with the voltage value exactly at the first Shapiro step, demonstrating the rectification from the *rf*-microwave to a quantized voltage. At $I_{dc} = 0$, the $V_{dc}$ output exhibits opposite Shapiro steps when the magnetic field is reversed. The modulation of the *rf* rectification effect by the magnetic field is further investigated with the $V_{dc}$ map plotted as a function of $I_{dc}$ and $B_z$ under a reduced *rf* power of 2.6 dBm (see Extended Data Fig. 7e). This map clearly shows a nonzero $V_{dc}$ at $I_{dc} = 0$ and a sign reversal when the magnetic field direction is flipped. The polarity reversal of the *rf* rectification through magnetic field flipping is further demonstrated by alternating between ±2.5 mT magnetic fields at $I_{dc} = 0$, as shown in Fig. 3e, where the output $V_{dc}$ continuously toggles between the "-h*f*/2e" and "h*f*/2e" states. Moreover, these quantized output voltages under *rf* irradiation pulses show

11 / 39plane magnetic field and *rf* power at $I_{dc} = 0$ and $f = 2$ GHz. A finite $V_{dc}$ is clearly observed under the application of a magnetic field. Moreover, as the *rf* power increases, a finite $V_{dc}$ can emerge even in the absence of both a magnetic field and a DC current bias, as indicated by the dashed line at $B_z = 0$ in Fig. 3a.

This $V_{dc}$ generated at $I_{dc} = 0$ is further confirmed through measurements of the AC Josephson effect under $B_z = \pm 2.5$ mT. As shown in Figs. 3b and 3c, the $dV/dI$ maps as a function of *rf* power and $I_{dc}$ reveal a pronounced asymmetry around zero $I_{dc}$ over the entire *rf* power range. The yellow dashed lines at $I_{dc} = 0$ in Figs. 3b and 3c indicate the emergence of nonzero DC voltages at specific *rf* powers (Extended Data Fig. 7). The typical $I - V$ characteristics at *rf* power 4.9 dBm are depicted in Fig. 3d, where the zeroth voltage step is fully positioned on one side of $I_{dc} = 0$. Clearly, the $V_{dc}$ exhibits nonzero output at $I_{dc} = 0$, with the voltage value exactly at the first Shapiro step, demonstrating the rectification from the *rf*-microwave to a quantized voltage. At $I_{dc} = 0$, the $V_{dc}$ output exhibits opposite Shapiro steps when the magnetic field is reversed. The modulation of the *rf* rectification effect by the magnetic field is further investigated with the $V_{dc}$ map plotted as a function of $I_{dc}$ and $B_z$ under a reduced *rf* power of 2.6 dBm (see Extended Data Fig. 7e). This map clearly shows a nonzero $V_{dc}$ at $I_{dc} = 0$ and a sign reversal when the magnetic field direction is flipped. The polarity reversal of the *rf* rectification through magnetic field flipping is further demonstrated by alternating between ±2.5 mT magnetic fields at $I_{dc} = 0$, as shown in Fig. 3e, where the output $V_{dc}$ continuously toggles between the "-h*f*/2e" and "h*f*/2e" states. Moreover, these quantized output voltages under *rf* irradiation pulses show consistent and repeatable on/off behavior (see Extended Data Figs. 7f and 7g). This rectification effect was further tested across frequencies from 0.2 to 4 GHz, covering the 2.4 GHz Wi-Fi band. As shown in Fig. 3f, the output $V_{dc}$ exhibits a linear relationship with frequency $f$, and all values fall on the $hf/2e$ plateau. The quantized DC voltage plateau at each frequency is obtained by a specific *rf* power. Intriguingly, besides the emergence of first Shapiro step at selected *rf* powers, the $V_{dc}$ output at $I_{dc}$ = 0 can also fall on other quantized voltage steps by varying the *rf* power (see Extended Data Fig. 8). At low *rf* power, $V_{dc}$ exhibits fractional voltage steps at $\frac{m}{9}\frac{hf}{2e}$, with the integer *m* incrementing sequentially as the power increases. In the high *rf* power regime, $V_{dc}$ shows discrete integer Shapiro steps at $n\frac{hf}{2e}$, where *n* also increases with *rf* power.

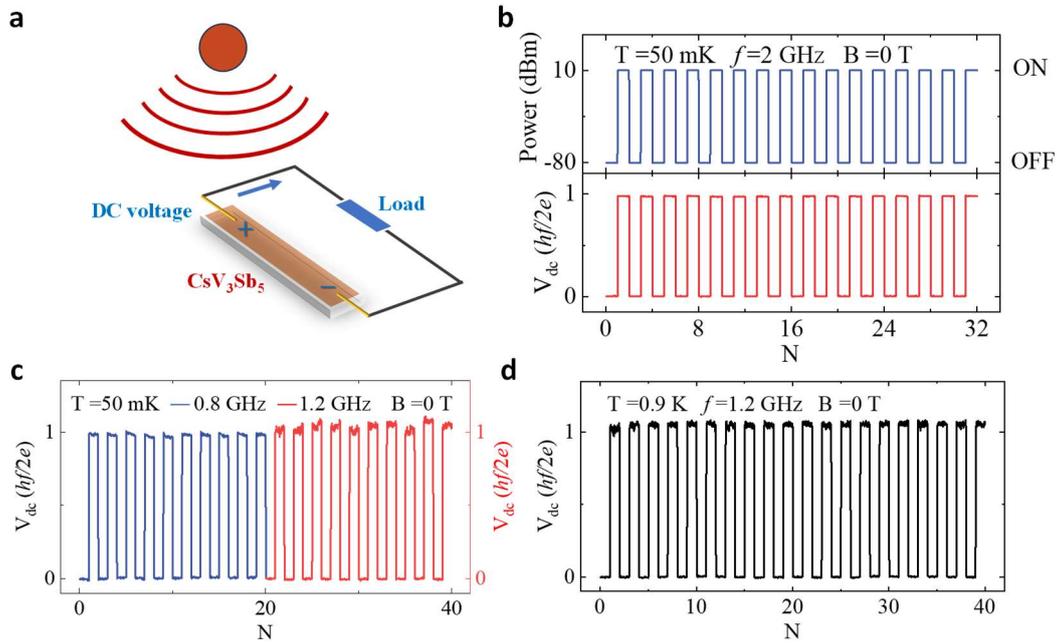

**Fig. 4: Quantized *rf* rectification in CsV$_3$Sb$_5$ without magnetic field. a**, Schematic of a rectifier based on the Josephson diode in monolithic CsV$_3$Sb$_5$ film. **b-d**, Quantized rectification under microwave irradiation at **b**, $f = 2$ GHz and power of 10.1 dBm; **c**,

0.8 GHz (14.3 dBm) and 1.2 GHz (11.4 dBm); **d,** $f = 1.2$ GHz and power of 11.5 dBm under a higher temperature of 0.9 K.

Furthermore, we construct the wireless rectifiers based on the Josephson diode in CsV$_3$Sb$_5$, as illustrated in Fig. 4a. At $B_z = 0$ T and $f = 2$ GHz, DC voltage on/off states are achieved under *rf* power pulses of 10.1 dBm, toggling between zero and the Shapiro step of $hf/2e$, as shown in Fig. 4b, demonstrating the quantized rectification effect. Figure 4c presents the quantized rectification measured at different frequencies, with selected *rf* power pulses of 14.3 dBm (0.8 GHz) and 11.4 dBm (1.2 GHz), respectively. Remarkably, the quantized rectification effect persists at a higher temperature of 0.9 K (Fig. 4d). Even at an elevated temperature of 1.6 K and a frequency of 4 GHz, the rectification effect remains detectable, although the output $V_{dc}$ deviates from the quantized Shapiro steps (see Extended Data Fig. 9). These findings demonstrate the feasibility of a robust radio-frequency rectifier based on CsV$_3$Sb$_5$ without magnetic field.

**Perspectives**

The observed supercurrent Fraunhofer interference patterns and Shapiro steps provide strong evidence for Josephson junctions in the kagome superconductor CsV$_3$Sb$_5$, consistent with theoretical predictions of superconducting domains due to multiple order parameters. Josephson junctions formed within the single material typically benefit from a high interface transmission, which may be a key factor enabling the robust AC Josephson effect that can persist near T$_c$. Moreover, the Josephson diode effect with unidirectional superconductivity is observed under the microwave radiation. Taking advantage of this Josephson diode effect, a wireless rectifier is realized in $CsV_3Sb_5$ that converts *rf* signals into quantized DC voltage without the need for a magnetic field or DC current source. This showcases significant potential for applications in non-dissipative superconducting circuits, wireless power, wireless charging, quantum measurement and control, and voltage standards, etc. Furthermore, this high-quality Josephson junction based on monolithic $CsV_3Sb_5$ sample can serve as a foundational building block for superconducting qubits, enhancing controllability and fidelity, and advancing the development of superconducting quantum computing[64–66].

**Online content**

Any methods, additional references, Nature Portfolio reporting summaries, source data, extended data, supplementary information, acknowledgements, peer review information; details of author contributions and competing interests; and statements of data and code availability are available at ~.

**Reference**

1.  Ando, F. et al. Observation of superconducting diode effect. *Nature* **584**, 373–376 (2020).
2.  Lin, J.-X. et al. Zero-field superconducting diode effect in small-twist-angle trilayer graphene. *Nat. Phys.* **18**, 1221–1227 (2022).
3.  Daido, A., Ikeda, Y. & Yanase, Y. Intrinsic Superconducting Diode Effect. *Phys. Rev. Lett.* **128**, 037001 (2022).
4.  Nadeem, M., Fuhrer, M. S. & Wang, X. The superconducting diode effect. *Nat. Rev. Phys.* **5**, 558–577 (2023).
5.  N. F. Q. Yuan, & L. Fu. Supercurrent diode effect and finite-momentum superconductors, *Proc.*

*Natl. Acad. Sci.* **119,** e2119548119 (2022).

6. Wakatsuki, R. et al. Nonreciprocal charge transport in noncentrosymmetric superconductors. *Sci. Adv.* **3**, e1602390 (2017).

7. Narita, H. et al. Field-free superconducting diode effect in noncentrosymmetric superconductor/ferromagnet multilayers. *Nat. Nanotechnol.* **17**, 823–828 (2022).

8. Ilić, S. & Bergeret, F. S. Theory of the Supercurrent Diode Effect in Rashba Superconductors with Arbitrary Disorder. *Phys. Rev. Lett.* **128**, 177001 (2022).

9. Ideue, T. & Iwasa, Y. One-way supercurrent achieved in an electrically polar film. *Nature,* **584**, 349-350 (2020).

10. Hou, Y. et al. Ubiquitous Superconducting Diode Effect in Superconductor Thin Films. *Phys. Rev. Lett.* **131**, 027001 (2023).

11. Miyasaka, Y. et al. Observation of nonreciprocal superconducting critical field. *Appl. Phys. Express* **14**, 073003 (2021).

12. Margarita Davydova et al. Universal Josephson diode effect. *Sci. Adv.* **8**, eabo0309 (2022).

13. Pal, B. et al. Josephson diode effect from Cooper pair momentum in a topological semimetal. *Nat. Phys.* **18**, 1228-1233 (2022).

14. Misaki, K. & Nagaosa, N. Theory of the nonreciprocal Josephson effect. *Phys. Rev. B* **103**, 245302 (2021).

15. Hu, J., Wu, C. & Dai, X. Proposed Design of a Josephson Diode. *Phys. Rev. Lett.* **99**, 067004 (2007).

16. Wu, H. et al. The field-free Josephson diode in a van der Waals heterostructure. *Nature* **604**, 653–656 (2022).

17. Baumgartner, C. et al. Supercurrent rectification and magnetochiral effects in symmetric Josephson junctions. *Nat. Nanotechnol.* **17**, 39–44 (2022).

18. Zhang, Y., Gu, Y., Li, P., Hu, J. & Jiang, K. General Theory of Josephson Diodes. *Phys. Rev. X* **12**, 041013 (2022).

19. Jeon, K.-R. et al. Zero-field polarity-reversible Josephson supercurrent diodes enabled by a proximity-magnetized Pt barrier. *Nat. Mater.* **21**, 1008–1013 (2022).

20. Seoane Souto, R. et al. Tuning the Josephson diode response with an ac current. *Phys. Rev.*

*Research* **6**, L022002 (2024).

21. Lotfizadeh, N. et al. Superconducting diode effect sign change in epitaxial Al-InAs Josephson junctions. *Commun. Phys.* **7**, 120 (2024).

22. Gupta, M. et al. Gate-tunable superconducting diode effect in a three-terminal Josephson device. *Nat. Commun.* **14**, 3078 (2023).

23. Ortiz, B. R. et al. New kagome prototype materials: discovery of $KV_3Sb_5$, $RbV_3Sb_5$, and $CsV_3Sb_5$. *Phys. Rev. Materials* **3**, 094407 (2019).

24. Ortiz, B. R. et al. $CsV_3Sb_5$: A $Z_2$ Topological Kagome Metal with a Superconducting Ground State. *Phys. Rev. Lett.* **125**, 247002 (2020).

25. Ortiz, B. R. et al. Superconductivity in the $Z_2$ kagome metal $KV_3Sb_5$. *Phys. Rev. Materials* **5**, 034801 (2021).

26. Yin, Q. et al. Superconductivity and Normal-State Properties of Kagome Metal $RbV_3Sb_5$ Single Crystals. *Chinese Phys. Lett.* **38**, 037403 (2021).

27. Hu, Y. et al. Rich nature of Van Hove singularities in Kagome superconductor Cs $V_3Sb_5$. *Nat. Commun.* **13**, 2220 (2022).

28. Kiesel, M. L., Platt, C. & Thomale, R. Unconventional Fermi Surface Instabilities in the Kagome Hubbard Model. *Phys. Rev. Lett.* **110**, 126405 (2013).

29. Kang, M. et al. Twofold van Hove singularity and origin of charge order in topological kagome superconductor $CsV_3Sb_5$. *Nat. Phys.* **18**, 301–308 (2022).

30. Zhao, H. et al. Cascade of correlated electron states in the kagome superconductor $CsV_3Sb_5$. *Nature* **599**, 216–221 (2021).

31. Li, H. et al. Observation of Unconventional Charge Density Wave without Acoustic Phonon Anomaly in Kagome Superconductors $AV_3Sb_5$ (A = Rb, Cs). *Phys. Rev. X* **11**, 031050 (2021).

32. Zheng, L. et al. Emergent charge order in pressurized kagome superconductor $CsV_3Sb_5$. *Nature* **611**, 682–687 (2022).

33. Tan, H., Liu, Y., Wang, Z. & Yan, B. Charge Density Waves and Electronic Properties of Superconducting Kagome Metals. *Phys. Rev. Lett.* **127**, 046401 (2021).

34. Chen, H. et al. Roton pair density wave in a strong-coupling kagome superconductor. *Nature* **599**, 222–228 (2021).

35. Jiang, K., Zhang, Y., Zhou, S. & Wang, Z. Chiral Spin Density Wave Order on the Frustrated Honeycomb and Bilayer Triangle Lattice Hubbard Model at Half-Filling. *Phys. Rev. Lett.* **114**, 216402 (2015).

36. Nie, L. et al. Charge-density-wave-driven electronic nematicity in a kagome superconductor. *Nature* **604**, 59–64 (2022).

37. Xu, Y. et al. Three-state nematicity and magneto-optical Kerr effect in the charge density waves in kagome superconductors. *Nat. Phys.* **18**, 1470–1475 (2022).

38. Yang, S.-Y. et al. Giant, unconventional anomalous Hall effect in the metallic frustrated magnet candidate, $KV_3Sb_5$. *Sci. Adv.* **6**, eabb6003 (2020).

39. Mielke, C. et al. Time-reversal symmetry-breaking charge order in a kagome superconductor. *Nature* **602**, 245–250 (2022).

40. Guo, C. et al. Switchable chiral transport in charge-ordered kagome metal $CsV_3Sb_5$. *Nature* **611**, 461–466 (2022).

41. Ming, F. et al. Evidence for chiral superconductivity on a silicon surface. *Nat. Phys.* **19**, 500–506 (2023).

42. Zhao, C. C. et al. Nodal superconductivity and superconducting domes in the topological Kagome metal CsV3Sb5. Preprint at https://arxiv.org/abs/2102.08356 (2021).

43. Deng, H. et al. Chiral kagome superconductivity modulations with residual Fermi arcs. *Nature* **632**, 775–781 (2024).

44. Wu, X. et al. Nature of Unconventional Pairing in the Kagome Superconductors $AV_3Sb_5$ ( A= K, Rb, Cs). *Phys. Rev. Lett.* **127**, 177001 (2021).

45. Zhou, S. & Wang, Z. Chern Fermi pocket, topological pair density wave, and charge-4e and charge-6e superconductivity in kagomé superconductors. *Nat. Commun.* **13**, 7288 (2022).

46. Saykin, D. R. et al. High Resolution Polar Kerr Effect Studies of $CsV_3Sb_5$: Tests for Time-Reversal Symmetry Breaking below the Charge-Order Transition. *Phys. Rev. Lett.* **131**, 016901 (2023).

47. Zhong, Y. et al. Nodeless electron pairing in $CsV_3Sb_5$-derived kagome superconductors. *Nature* **617**, 488–492 (2023).

48. Mu, C. et al. S-Wave Superconductivity in Kagome Metal $CsV_3Sb_5$ Revealed by $^{121/123}$Sb NQR

and[51] V NMR Measurements. *Chinese Phys. Lett.* **38**, 077402 (2021).

49. Le, T. et al. Superconducting diode effect and interference patterns in kagome Cs V$_3$Sb$_5$. *Nature* **630**, 64–69 (2024).

50. Christensen, M. H. et al. Loop currents in AV$_3$Sb$_5$ kagome metals: Multipolar and toroidal magnetic orders. *Phys. Rev. B* **106**, 144504 (2022).

51. Feng, X. et al. Low-energy effective theory and symmetry classification of flux phases on the kagome lattice. *Phys. Rev. B* **104**, 165136 (2021).

52. Shapiro, S. Josephson Currents in Superconducting Tunneling: The Effect of Microwaves and Other Observations. *Phys. Rev. Lett.* **11**, 80–82 (1963).

53. Russer, P. Influence of Microwave Radiation on Current-Voltage Characteristic of Superconducting Weak Links. *J. Appl. Phys.* **43**, 2008-2010 (1972).

54. Devoret, M. H. & Schoelkopf, R. J. Superconducting Circuits for Quantum Information: An Outlook. *Quantum Information Processing* **339**, (2013).

55. Rüfenacht, A., Flowers-Jacobs, N. E. & Benz, S. P. Impact of the latest generation of Josephson voltage standards in ac and dc electric metrology. *Metrologia* **55**, S152–S173 (2018).

56. Howe, L. et al. Digital Control of a Superconducting Qubit Using a Josephson Pulse Generator at 3 K. *PRX Quantum* **3**, 010350 (2022).

57. F. Kidwingira, J. D. Strand, D. J. Van Harlingen, and Y. Maeno, Dynamical Superconducting Order Parameter Domains in Sr$_2$RuO$_4$, *Science* **314**,1267-1271(2006).

58. Xiang, Y. et al. Twofold symmetry of c-axis resistivity in topological kagome superconductor CsV$_3$Sb$_5$ with in-plane rotating magnetic field. *Nat. Commun.* **12**, 6727 (2021).

59. Likharev, K. K. Superconducting weak links. *Rev. Mod. Phys.* **51**, 101–159 (1979).

60. M. S. Hossain et al. Unconventional gapping behavior in a Kagome superconductor, *Nat. Phys.* **21**, 556–563 (2025).

61. Y. Tanaka et al. Domains in multiband superconductors, *Physica C* **471**, 747 (2011).

62. Su, H. et al. Microwave-Assisted Unidirectional Superconductivity in Al-InAs Nanowire-Al Junctions under Magnetic Fields. *Phys. Rev. Lett.* **133**, 087001 (2024).

63. Daido, A. & Yanase, Y. Unidirectional Superconductivity and Diode Effect Induced by Dissipation. Preprint at http://arxiv.org/abs/2310.02539 (2023).

64. Linder, J. & Robinson, J. W. A. Superconducting spintronics. *Nature Phys.* **11**, 307–315 (2015).

65. Clarke, J. & Wilhelm, F. K. Superconducting quantum bits. *Nature* **453**, 1031–1042 (2008).

66. Huang, H.-L., Wu, D., Fan, D. & Zhu, X. Superconducting quantum computing: a review. *Sci. China Inf. Sci.* **63**, 180501 (2020).

## Methods

### Device fabrication

The CsV$_3$Sb$_5$ strips were mechanically exfoliated from a bulk crystal to a high-resistance silicon substrate. The whole exfoliating process was done in an argon-filled glove box with $O_2$ and $H_2O$ content below 0.01 parts per million to avoid sample degeneration. The thicknesses of the selected CsV$_3$Sb$_5$ films are around 100~200 nm, and the widths of which are around 1~2.5 μm. Following an Ar etching procedure to eliminate the surface oxidized layer, the Ti/Au electrodes (5/145 nm thick) were fabricated by electron beam evaporation for transport measurement.

### Transport measurements

The transport measurements were carried out in a dilution refrigerator with a variable temperature insert and a superconducting magnet. The standard lock-in techniques (Stanford Research Systems 830) combined with a 1 MΩ buffer resistor were used to supply a small AC current $I_{ac}$ and collect the signals of differential resistance $dV/dI = V_{ac}/I_{ac}$ with frequency ω. The frequency $\omega$ is 17.777 Hz unless otherwise stated. The direct current source meter Keithley 2400 was used to inject direct current $I_{dc}$ and the nano-volt meter Keithley 2182 was used to measure the DC voltage $V_{dc}$. The differential resistance $dV/dI$ in the positive ($I_{dc}^+$) and negative ($I_{dc}^-$) current regions were obtained by sweeping the direct current $I_{dc}$ from 0 to $I_{max}^+$ and $I_{max}^-$, respectively. The measurements were performed by standard four-terminal methods.

**Microwave power applied on the sample**

The AC Josephson effect was measured by coupling the microwave to the device. The microwave power level mentioned in the manuscript refers to the output power of the external microwave generator, rather than the actual power dissipated by the sample. The microwave transmission line within the dilution refrigerator is equipped with four attenuators, providing a total attenuation of 28 dB. This attenuation helps prevent noise leakage into the sample, which could otherwise lead to unwanted heating. Additionally, due to the impedance mismatch between the sample and the final part of the transmission line with the 50 Ω line, the microwave power coupled to the sample is further reduced. Usually, for a 0 dBm (1 mW) output from the generator, the microwave power coupling to the sample is approximately −50 dBm (i.e., $10^{-5}$ mW). Due to the impedance mismatch between the sample and the microwave line, the current bias condition can be assumed for the microwave signal.

**Relationship between superconducting transition peaks and Shapiro steps**

Various sets of Shapiro steps are observed in the AC Josephson effect (Fig. 2a). As shown in Extended Data Fig. 4a, we identify three distinct sets of Shapiro steps which may be attributed to different Josephson junctions, labeled $JJ_1$, $JJ_2$, and $JJ_3$.

- $JJ_1$ (Fractional Steps): Dominates at low direct current $I_{dc}$, exhibiting fractional Shapiro steps with step height $\frac{1}{9}\frac{hf}{2e}$, as demonstrated in Figs. 2c and 2d Remarkably, these fractional steps persist even at higher microwave powers (Extended Data Fig. 4b), although they require extremely low $I_{dc}$.

- JJ$_2$ (Integer Steps): Emerges at intermediate $I_{dc}$, with $n = 1, 2, \ldots$. As $I_{dc}$ increases, the Shapiro steps transition from fractional to integer values with a step height of $\frac{hf}{2e}$, as shown in Fig. 2e.

- JJ$_3$ (Higher-Order Steps): Activates at high $I_{dc}$, showing $\Delta V = \frac{hf}{e}$. At relatively high $I_{dc}$ and high power, as $I_{dc}$ increases (Fig. 2f) or as *rf* power increases (Extended Data Fig. 4c), the Shapiro steps evolve further, from $\frac{hf}{2e}$ step height to $\frac{hf}{e}$ step height. Notably, this $\frac{hf}{e}$ step height might be the result of two junctions in series (JJ$_2$+JJ$_3$': $\frac{hf}{2e} + \frac{hf}{2e}$).

Multiple transitions can be observed in the $I - V$ curves (Fig. 1e) and the superconducting interference patterns (Fig. 1f). The first three critical currents of the transition peaks are labelled as $I_c$, $I_c'$, and $I_c''$ (Extended Data Fig. 5a). The critical currents of JJ$_1$, JJ$_2$, and JJ$_3$ (extracted from Shapiro step thresholds in Extended Data Fig. 4a) align precisely with $I_c$, $I_c'$, and $I_c''$, as shown in Extended Data Fig. 5b. This direct correspondence of the critical currents between the Shapiro steps and the transition peaks provides evidence that the various sets of Shapiro steps, JJ$_1$, JJ$_2$ and JJ$_3$, arise from different superconducting transitions.

Extended Data Figure 5c depicts the $I - V$ curves under *rf* irradiation applied with different *rf* powers. At low $I_{dc}$, fractional Shapiro steps (JJ$_1$) emerge near $I_c$. As $I_{dc}$ approaches $I_c'$, integer steps (JJ$_2$) dominate. This evolution in the Shapiro steps suggest that the fractional and integer steps originate from distinct transitions. In addition, we examine the superconducting interferences under *rf* irradiation. As shown in Extended Data Fig. 5d, three distinct sets of interference patterns are observed (denoted by black arrows), directly corresponding to specific different Josephson junctions (JJ$_1$, JJ$_2$, and JJ$_3$). This observation further confirms that the different sets of Shapiro steps stem from the multiple peak superconducting transitions, suggesting the existence of multiple Josephson junctions.

**Quantized rectification voltage**

We now try to provide a physical interpretation for the quantized rectification voltage occurring in a Josephson diode. For simplicity, here we consider a Josephson diode with a current-phase relation (CPR) $I_s(\varphi) = I_{c+}sin\varphi$ for $I_s > 0$ and $I_s(\varphi) = I_{c-}sin\varphi$ for $I_s < 0$, where $I_{c+} \neq |I_{c-}|$. Such CPR can well capture the non-reciprocal characteristic of critical supercurrent observed in the CsV$_3$Sb$_5$ junction. In the following, we would adopt the CPR to elaborate on the quantized rectification voltage in CsV$_3$Sb$_5$. As for the exact form of CPR in CsV$_3$Sb$_5$ and its underlying mechanism, they cannot be conclusively determined from the current data and require further experimental and theoretical investigations in the future, which lie beyond the scope of this work.

Since the sample resistance is much lower than the impedance of the microwave source, the driving source acts as a current source. The current biased Josephson junction can be understood in terms of a tilted washboard potential against a phase particle $\varphi$ (Ref.[67]):

$$U(\varphi) = E_J(1 - cos\varphi) - \frac{\hbar I\varphi}{2e}, \tag{1}$$

where $U$ is the potential, $E_J = \frac{\hbar I_c}{2e}$ is the Josephson energy, $\varphi$ is the phase difference across the junction, and $I$ is the bias current.

For the Josephson diode $I_{c+} \neq |I_{c-}|$, we can rewrite the potential $U$ as:

$$U(\varphi) = \begin{cases} E_{J+}(1 - cos\varphi) - \dfrac{\hbar I \varphi}{2e} + 2n(E_{J+} - E_{J-}), for\ \varphi \in [2n\pi,\ (2n+1)\pi] \\ E_{J-}(1 - cos\varphi) - \dfrac{\hbar I \varphi}{2e} + 2n(E_{J+} - E_{J-}), for\ \varphi \in [(2n-1)\pi,\ 2n\pi] \end{cases}. \quad (2)$$

$$E_J = \begin{cases} E_{J+} = \dfrac{\hbar I_{c+}}{2e} \\ E_{J-} = \dfrac{\hbar |I_{c-}|}{2e} \end{cases}. \quad (3)$$

As shown in Extended Data Fig. 10a, it is clear that the potential is already tilted even without current while $E_{J+} \neq E_{J-}$. In order to understand dynamics of the phase driven by radio frequency current $I = I_{ac} sin2\pi ft$, we study the phase particle in the tilted washboard potential $U$ for one period of time $\Delta t = 1/f$ with $I_{c+} < I_{ac} < |I_{c-}|$ as displayed in Extended Data Fig. 10b. At $t = 0$, the current $I = 0$, the phase particle $\varphi$ is trapped in one of the potential wells. With time going from $t = 0$ to $t = \Delta t/4$, the current $I$ increases from $0$ to $I_{ac}$, and the washboard potential $U$ will be tilted more to the right. The phase particle will be able to roll down to the next lower potential well. With time going from $t = \Delta t/4$ to $t = \Delta t/2$, the slope of the tilted potential decreases, and the phase particle is trapped in a new potential well. With time going from $t = \Delta t/2$ to $t = 3\Delta t/4$, the washboard potential is tilted to the left. However, due to $I_{ac} < |I_{c-}|$, the phase particle will not be able to roll down back to the potential well on the left. With time going from $t = 3\Delta t/4$ to $t = \Delta t$, the washboard potential goes back to $I = 0$ shape and the phase particle remains trapped in the new potential well. Therefore, during this period of time $\Delta t$, we have phase slip $\Delta \varphi = 2\pi$ to the right.

According to Josephson effect, we have a quantized rectification voltage:

$$V = \frac{\hbar}{2e}\frac{d\varphi}{dt} = \frac{\hbar}{2e}\frac{\Delta\varphi}{\Delta t} = \frac{hf}{2e} = \Phi_0 f, \qquad (4)$$

where $\Phi_0 = h/2e$ is flux quantum. The quantized rectification voltage also can be understood as unidirectional periodic vortex motion across the junction while the overall charge tunneling through the junction is zero. Note that since phase and Cooper pairs number are conjugated variables, following Heisenberg uncertainty principle, once phase is known, Cooper pairs number is uncertain, however, the overall Cooper pairs tunneling through the junction on time average should be zero due to DC current is zero.

If the $I_{ac}$ is large enough, the phase slip can be $2m\pi$ ($m > 1$) for a period of time, where $m$ is a natural number, indicating the phase particle rolling down across $m$ potential wells. This will generate quantized rectification voltage $V = mhf/2e$ ($m > 1$). While $I_{ac}$ is even larger, phase slip will be $2(m-k)\pi$, where $k$ is a natural number, with phase slip $2m\pi$ on positive current and $-2k\pi$ on negative current for a period of time. Experimentally, we have indeed observed a variation of quantized rectification voltage as increasing the microwave power, as shown in Extended Data Figs. 7 and 8. As for the observed fractional rectification voltage, it may arise from the higher-harmonic component in CPR or measurement artifact in this device (See more discussions about the emergence of these fractional steps in Supplemental Note 5).

It is easy to see that with time period $\Delta t$ increasing, phase slip could increase while keeping the same amplitude $I_{ac}$, due to that there is more time for the phase particle rolling down to further potential well before trapped. However, the phase slip per period time $\Delta\varphi/\Delta t$ could be the same.

**Numerical simulation**

For a current driving Josephson junction, according to RCSJ model, the total current across the junction is

$$I = \frac{\hbar C}{2e}\frac{\partial^2\varphi}{\partial t^2} + \frac{\hbar}{2eR}\frac{\partial\varphi}{\partial t} + I_s(\varphi) = I_{dc} + I_{ac}sin2\pi ft, \tag{5}$$

where $C$ and $R$ are the shunted capacitance and resistance respectively. Since there is no hysteresis for differential resistance with current sweeping up and sweeping down (Extended Data Fig. 10d), the junction is in overdamping regime and the small shunted capacitance can be ignored. The equation can be written as

$$\frac{\hbar}{2e}\frac{\partial\varphi}{\partial t} + I_s(\varphi) = I_{dc} + I_{ac}sin2\pi ft. \tag{6}$$

For a Josephson diode with $I_{c+} \neq |I_{c-}|$, we have

$$\frac{\partial\varphi}{\partial\tau} = \begin{cases} i_{dc} + i_{ac}sin2\pi\Omega\tau - sin\varphi, for\ i_{dc} + i_{ac}sin2\pi\Omega\tau > 0 \\ i_{dc} + i_{ac}sin2\pi\Omega\tau - \lambda sin\varphi, for\ i_{dc} + i_{ac}sin2\pi\Omega\tau \leq 0 \end{cases}, \tag{7}$$

where $\tau = 2eI_{c+}Rt/\hbar$, $\lambda = I_{c-}/I_{c+}$, $i_{dc} = I_{dc}/I_{c+}$ and $i_{ac} = I_{ac}/I_{c+}$ is normalized current, and $\Omega = \hbar f/2eI_{c+}R$ is reduced frequency. Extended Data Figure 10e shows the numerical simulation according to the Eqs. (5-7) with $\lambda = 2$ and $\Omega = 0.04$. It can be seen that the differential resistance map is asymmetric with the $I_{dc} = 0$, which is consistent with the observation in Fig. 3b. This suggests that the *rf* rectification voltage under zero direct current should be the consequence of AC Josephson effect in a Josephson diode.

**Data availability**

All data needed to evaluate the conclusions in the paper are present in the paper. Additional data are available from the corresponding authors upon reasonable request. Source data are provided with this paper.

**Reference**

67. Tinkham, M. Introduction to superconductivity. 2nd edn 204-205 (McGraw-Hill, New York, 1996).

**Acknowledgements**

This work was supported by National Natural Science Foundation of China (Grant Nos. 62425401 and 62321004), and Innovation Program for Quantum Science and Technology (Grant No. 2021ZD0302403). C.Li acknowledged Dutch Research Council (NWO) for the financial support of the project SuperHOTS with file number VI.Vidi.203.047.

**Author contributions**

Z.-M.L. conceived and supervised the project. H.-X.L., X.L. and Q.Y. fabricated the devices. Z.-B.T. and J.-J.C. with the guidance of D.-P.Y. performed the transport measurements. J.-Z.F., X.-Y.L. and Y.-L.H. with the guidance of Z.-M.W. performed the SdH measurements. X.-M.M. conducted the STEM characterization. Z.-M.L., H.-X.L., X.-G.Y. and A.-Q.W. analyzed the data. Z.-M.L., H.-X.L., X.-G.Y., Z.-B.T., J.-

J.C, C.L. and A.-Q.W. wrote the manuscript. All authors discussed the results and commented on the manuscript.

## Competing interests

The authors declare no competing interests.

## Additional information

**Supplementary information** The online version contains supplementary material available at ~.

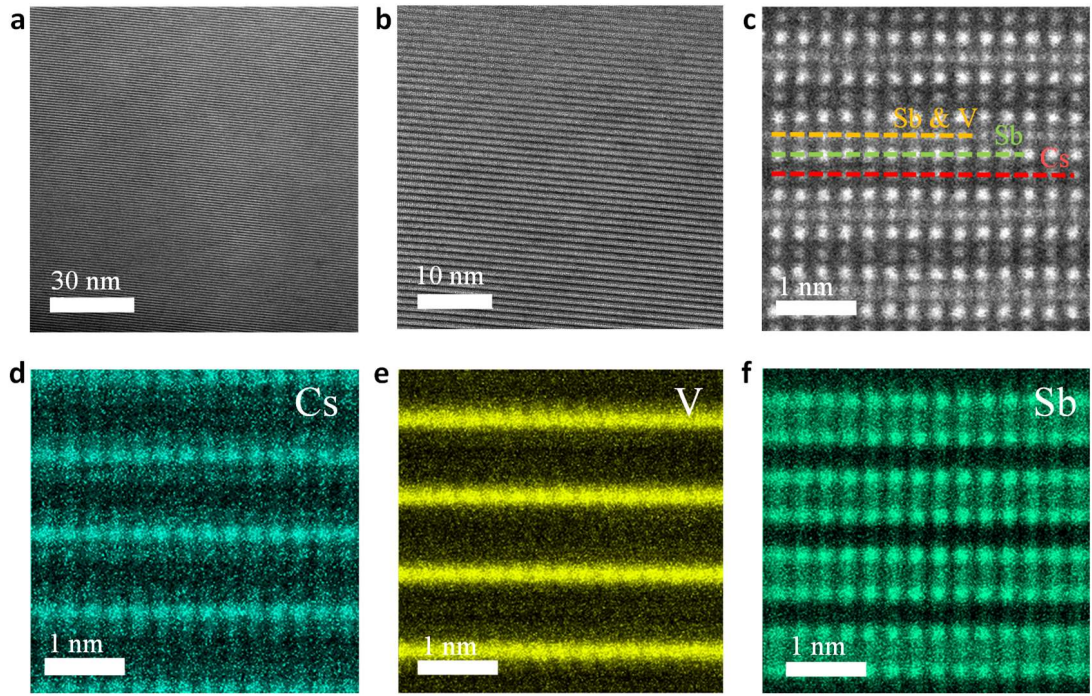

**Extended Data Fig. 1 | The scanning transmission electron microscopy (STEM) images of Device 2. a-c,** The typical STEM images (**a** and **b**) and the high angle annular dark field STEM (HAADF-STEM) image (**c**) captured within the sample present the uniform layered atomic structure of $CsV_3Sb_5$. **d-f**, Distribution maps of each element, Cs (**d**), V (**e**) and Sb (**f**) studied by an energy dispersive X-ray spectroscopy (EDX).

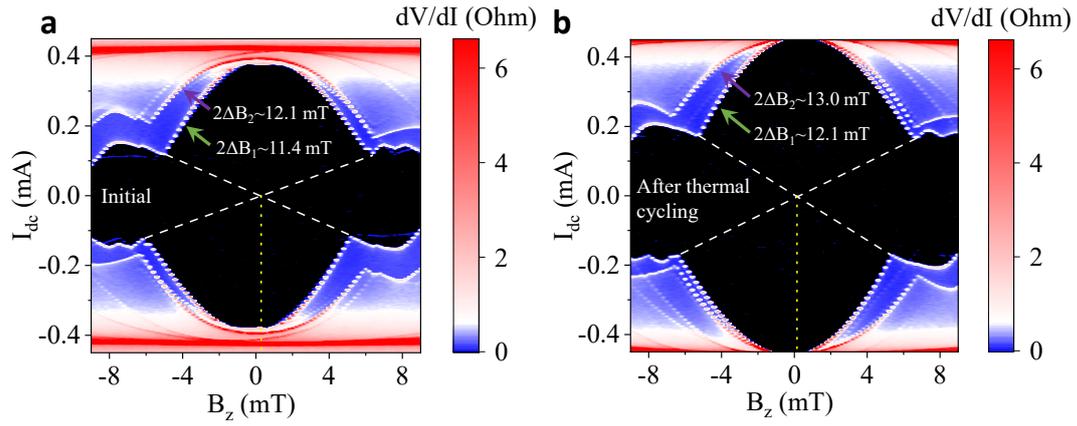

**Extended Data Fig. 2 | Thermal modulation of superconductivity interference patterns by using a local heater. a,** Initial color map of $dV/dI$ as a function of $I_{dc}$ and $B_z$. Note that the patterns in this figure and in Fig. 1f exhibit variations, as the measurements were conducted on different batches used for comparison. **b,** Color map of $dV/dI$ as a function of $I_{dc}$ and $B_z$ obtained after thermal cycling from the temperature slightly above $T_c$, which was conducted by using a local heater integrated near Device 1. The oscillation periods of both $I_c'$ (green arrow) and $I_c''$ (purple arrow) are changed after thermal cycles. It is also found that the critical currents $I_c'$ and $I_c''$ are increased after thermal cycles. All data were acquired between electrodes 2-3 of Device 1 at 50 mK.

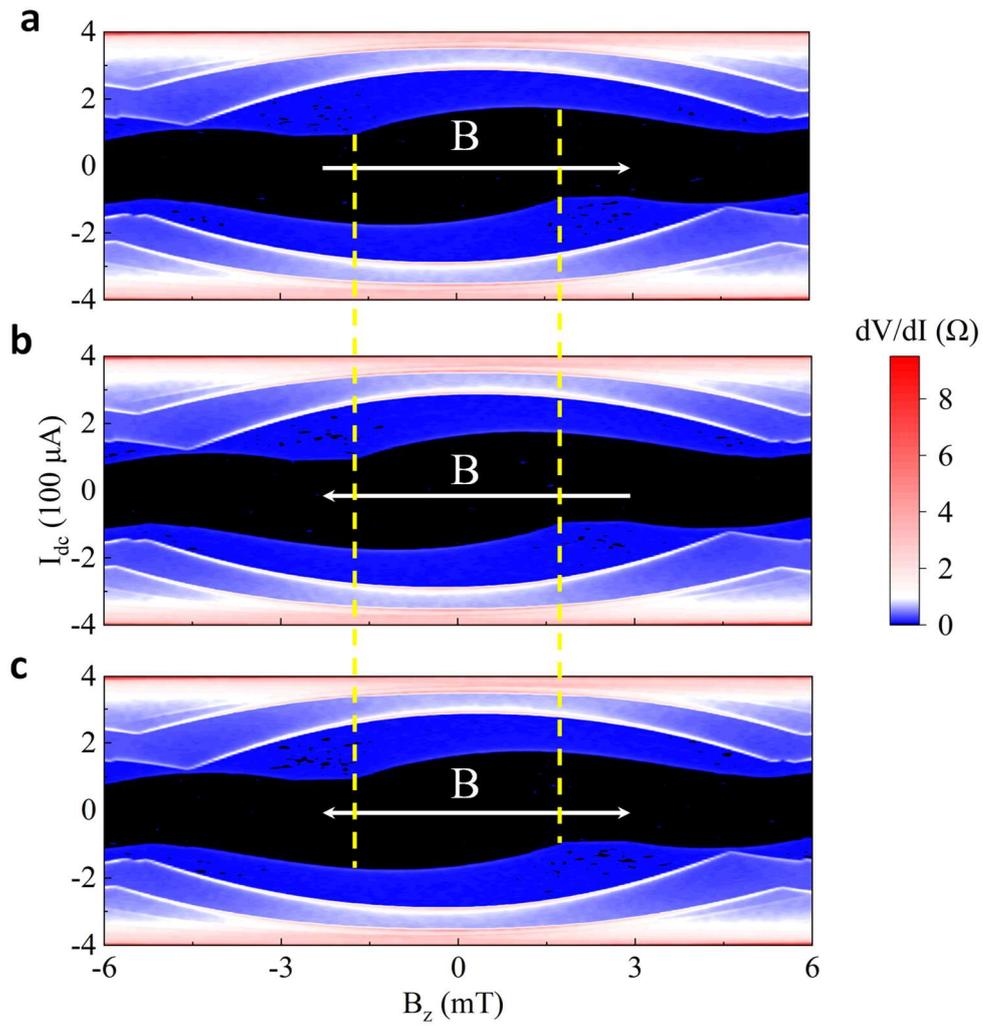

**Extended Data Fig. 3 | Superconductivity interference patterns obtained by different field-sweeping directions.** The mappings obtained by sweeping magnetic field $B_z$ from -6 to 6 mT (**a**), 6 to -6 mT (**b**), and 0 to 6 mT then to -6 mT (**e**), respectively. The vertical yellow dashed lines are guides to eyes. There is no visible phase shift in three superconductivity interference patterns measured under different field-sweeping directions, demonstrating the absence of trapped vortices in the measurement.

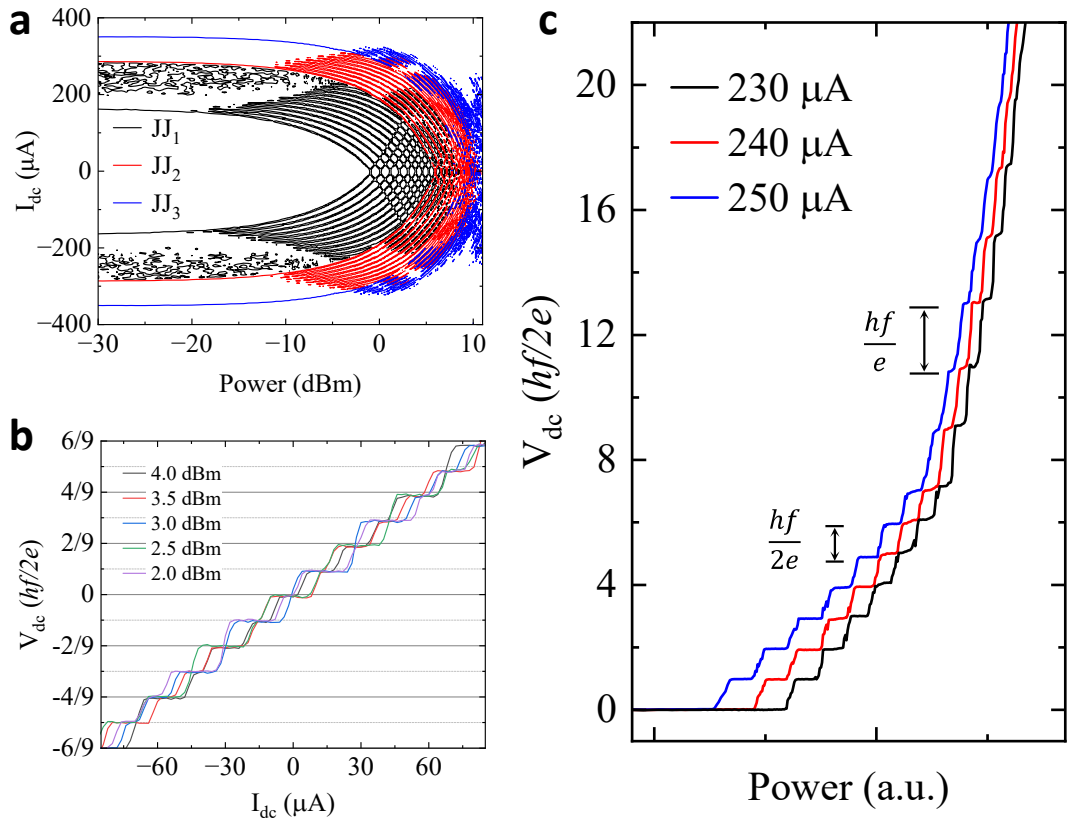

**Extended Data Fig. 4 | Multiple sets of Shapiro steps. a,** Three different sets of Shapiro steps, extracted from the contour lines in the Fig. 2a, which correspond to three different Josephson junctions JJ$_1$, JJ$_2$ and JJ$_3$. **b,** $I-V$ curves at low *rf* powers and low $I_{dc}$, presenting the fractional Shapiro steps. **c,** Power dependent DC voltage $V_{dc}$, presenting the evolution of Shapiro steps from *hf/2e* to *hf/e* at relatively large $I_{dc}$.

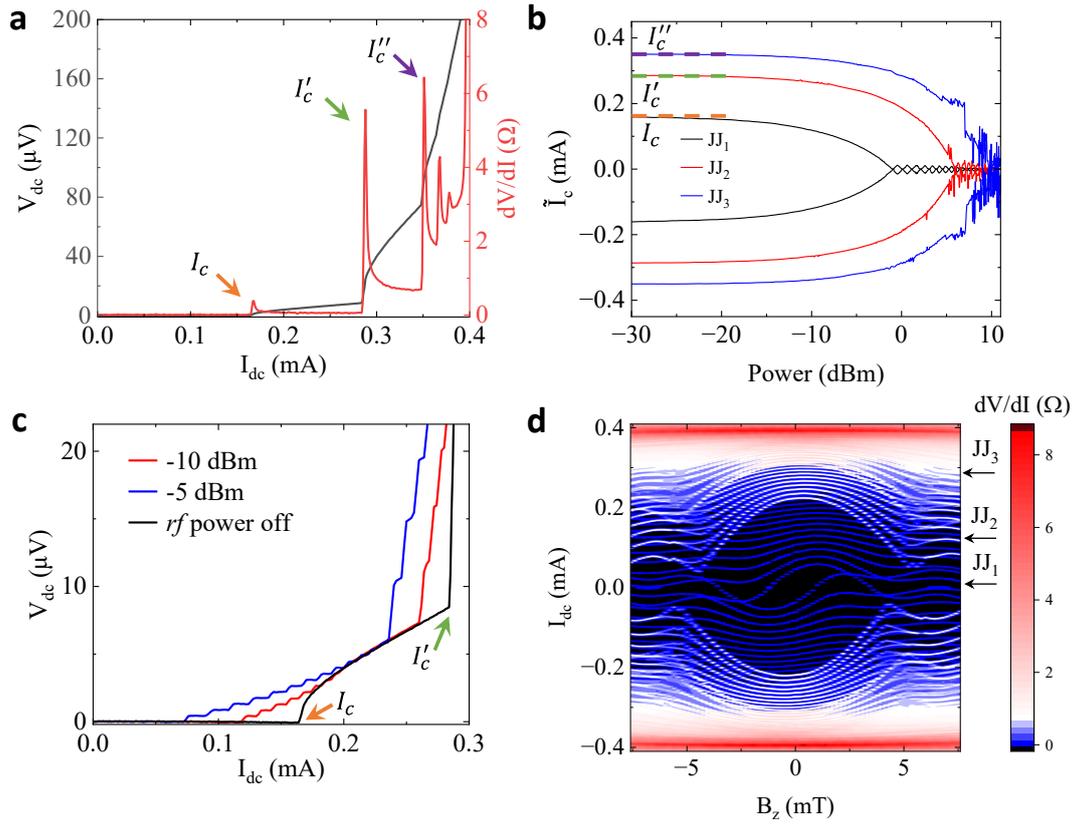

**Extended Data Fig. 5 | Relationship between transition peaks and Shapiro steps.** **a,** $dV/dI$ and $V_{dc}$ as the functions of $I_{dc}$, showing various superconducting transition peaks. **b,** Power dependent critical currents of JJ$_1$, JJ$_2$ and JJ$_3$. **c,** $I$-$V$ curves under *rf* irradiation with different *rf* powers. **d,** Color map of $dV/dI$ as a function of $I_{dc}$ and $B_z$ at $T = 50$ mK $f = 2$ GHz and P $= -2.6$ dBm, presenting multiple distinct sets of interference patterns.

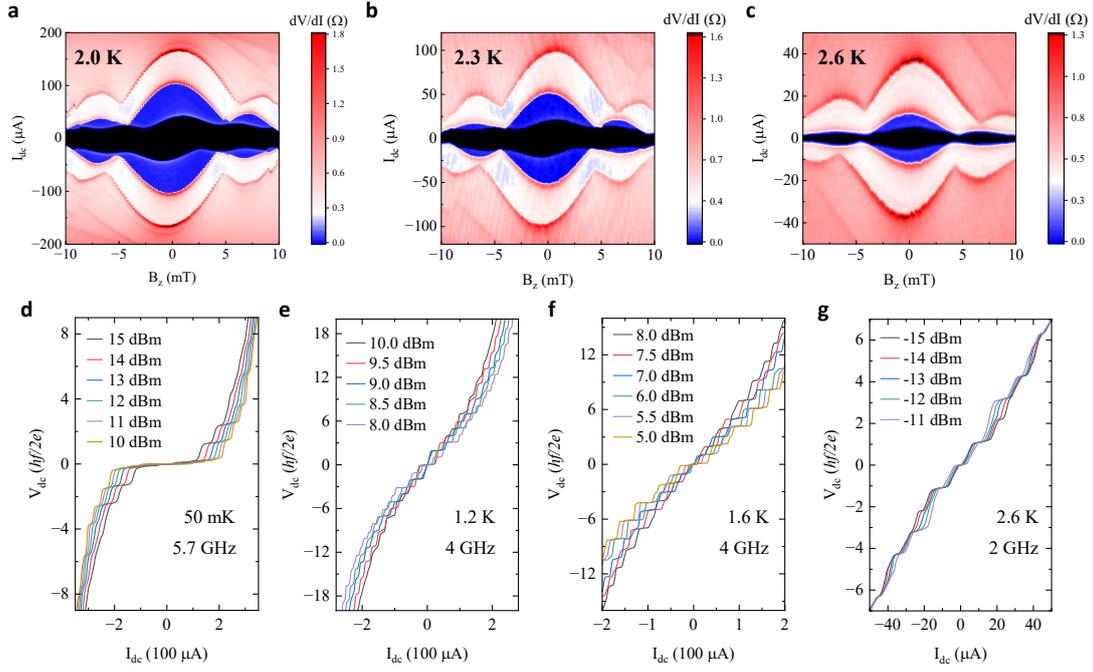

**Extended Data Fig. 6 | Temperature dependence of superconducting interference patterns and AC Josephson effect. a-c**, The differential resistance dV/dI map as a function of $I_{dc}$ and $B_z$ obtained at 2.0 K (**a**), 2.3 K (**b**) and 2.6 K (**c**), respectively, showing robust superconducting interference patterns against temperature. **d-g,** $I-V$ curves measured at different temperatures, with various *rf* powers applied. The integer Shapiro steps can be clearly observed, showing robustness against temperature.

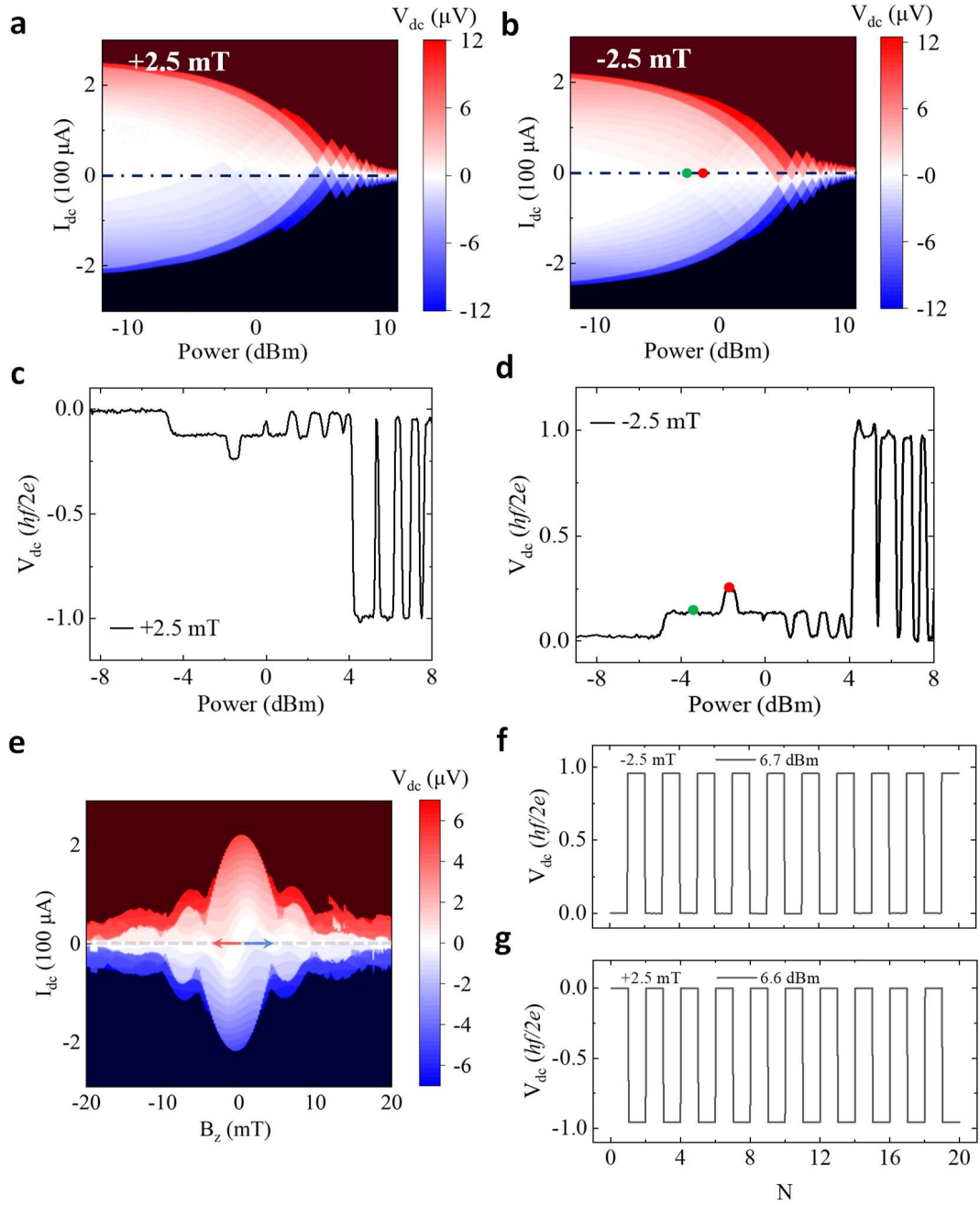

**Extended Data Fig. 7 | *rf* response modulated by magnetic field. a and b**, Mapping of $V_{dc}$ as a function of $I_{dc}$ and *rf* power with magnetic field of 2.5 mT (**a**) and -2.5 mT (**b**), **c and d**, respectively. The nonzero DC voltages emerge at specific *rf* powers without external current source, presenting quantized steps. The $V_{dc}$ output exhibits oppositive steps at positive magnetic field (**c**) and negative magnetic field (**d**). **e**, The DC voltage map as a function of $I_{dc}$ and $B_z$ is obtained at $T = 50$ mK and $f = 2$ GHz, with a *rf* power of 2.6 dBm. The nonzero DC voltage can be observed along

the zero current cut line (the gray dotted line), with the small magnetic fields applied. The polarity of output DC voltage is reversed when flipping the direction of magnetic field. **f** and **g**, Quantized rectification under magnetic fields. The DC voltage on/off switching can be realized by a rf power pulse of 6.7 dBm with an out of-plane magnetic field $B_z$ = -2.5 mT applied (**f**). When applying a magnetic field $B_z$ = 2.5 mT in the opposite direction, the similar rectification can be achieved by a *rf* power pulse of 6.6 dBm, but the polarity of the rectification is reversed (**g**).

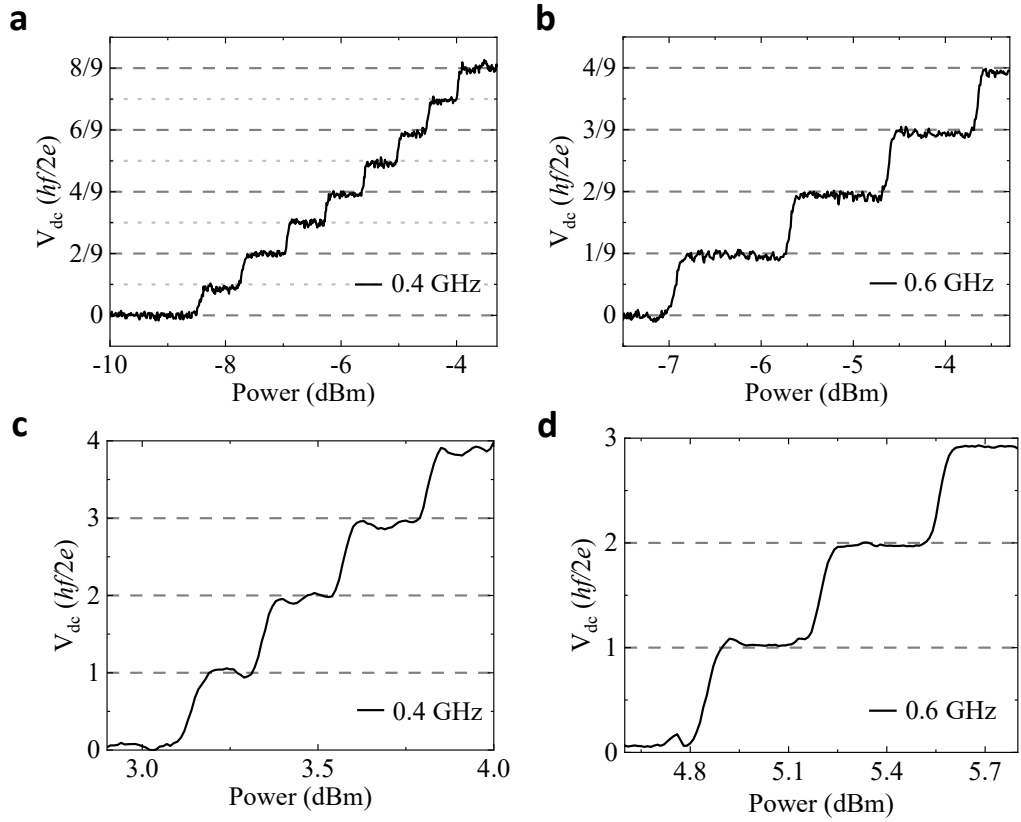

**Extended Data Fig. 8 | Power dependence of the output DC voltage.** The output DC voltage measured at $I_{dc}=0$, $B_z = -2.5$ mT with frequency $f$ of 0.4 GHz (**a** and **c**) and 0.6 GHz (**b** and **d**), respectively. At low *rf* power range, the $V_{dc}$ exhibits a series of fractional voltage steps (**a** and **b**), while the integer voltage steps emerge one by one at higher *rf* powers.

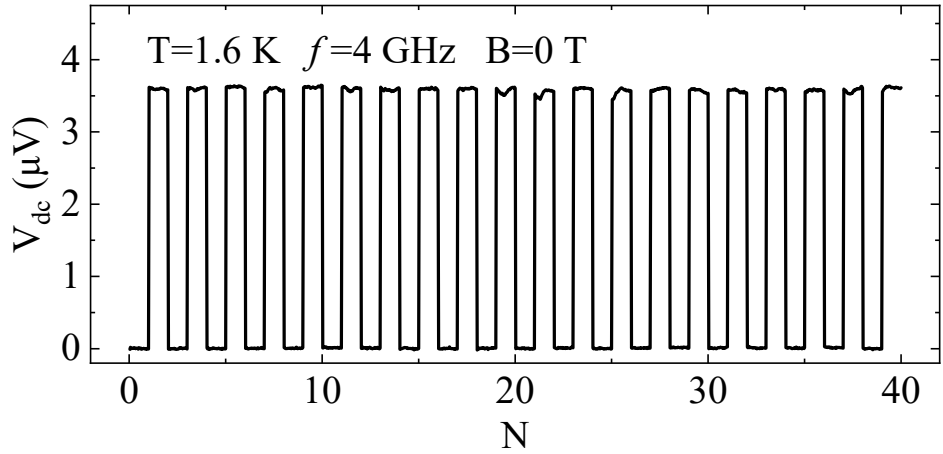

**Extended Data Fig. 9 | Rectification at a higher temperature and higher frequency.** At $T = 1.6$ K, $f = 4$ GHz and $B = 0$ T, DC voltage on/off states are achieved with a *rf* power pulses of 10.2 dBm.

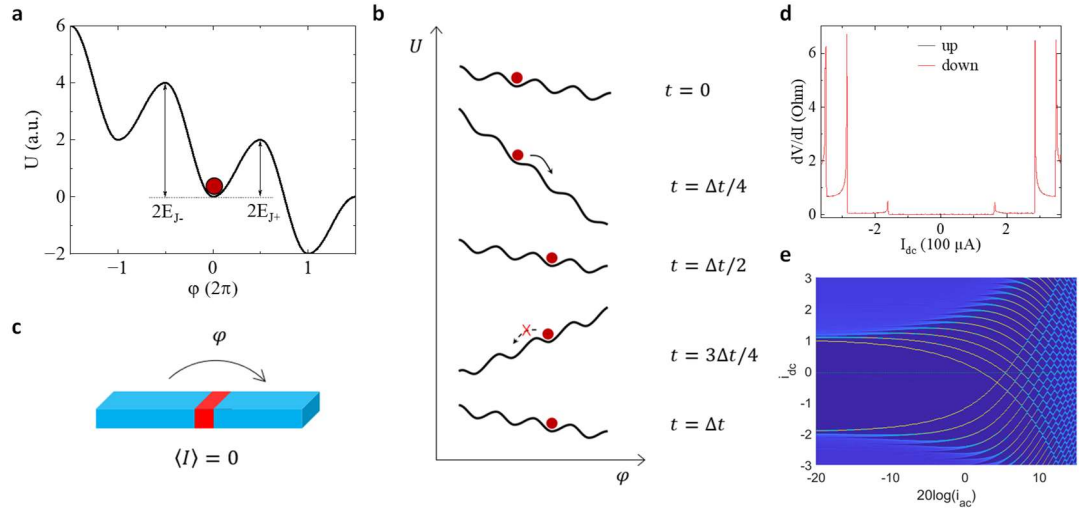

**Extended Data Fig. 10 | Quantized rectification voltage and AC Josephson effect of a Josephson diode. a,** The potential $U$ as a function of phase $\varphi$ with $E_{J+} \neq E_{J-}$ at zero bias current. **b,** Phase particle $\varphi$ evolution in a tilted washboard potential $U$ in a period of time $\Delta t$. **c,** Illustration of phase slip across the junction with zero overall charge transferred. **d,** Differential resistance shows no hysteresis with current sweeping up and sweeping down, indicating it is an overdamping junction. **e,** Numerical simulation of the differential resistance evolution with normalized current $i_{dc}$ and $20\log(i_{ac})$, with $\lambda = 2$ and $\Omega = 0.04$.